\documentclass[12pt,preprint]{aastex}
\usepackage{amsmath,amssymb}
\usepackage{natbib}
\usepackage{epsfig,graphics}
\usepackage{graphicx}
\usepackage{epstopdf} 
\usepackage{ulem}
\newcommand{\lsim}{\raise0.3ex\hbox{$<$}\kern-0.75em{\lower0.65ex\hbox{$\sim$}}}
\newcommand{\gsim}{\raise0.3ex\hbox{$>$}\kern-0.75em{\lower0.65ex\hbox{$\sim$}}}

\newcommand{\ds}{$\delta$-Scuti\ }
\begin{document}
%\slugcomment{Draft Version; not for circulation}
%-----------------------------------------------------------------------------%
\title{Spectroscopic Evidence for a Temperature Inversion in the Dayside Atmosphere of the Hot Jupiter WASP-33b}
%-----------------------------------------------------------------------------%

%------------------------------
%Author list - emulateapj2 style
%------------------------------
%\author{name\altaffilmark{1}}
%\altaffiltext{1}{affiliation}

%---------------------------
%Author list - journal style 
%---------------------------

\author{ Korey Haynes\altaffilmark{1,2,6}, Avi M. Mandell\altaffilmark{1}, Nikku Madhusudhan\altaffilmark{3}, Drake Deming\altaffilmark{4} , Heather Knutson\altaffilmark{5}
}
\altaffiltext{1}{Solar System Exploration Division, NASA Goddard Space Flight Center, Greenbelt, MD 20771, USA}
\altaffiltext{2}{School of Physics, Astronomy, and Computational Sciences, George Mason University, Fairfax, VA 22030, USA}
\altaffiltext{3}{Institute of Astronomy, University of Cambridge, Cambridge CB3 0HA, UK}
\altaffiltext{4}{Department of Astronomy, University of Maryland, College Park, MD 20742, USA}
\altaffiltext{5}{Division of Geological \& Planetary Sciences, California Institute of Technology, Pasadena, CA 91125, USA}
\altaffiltext{6}{Corresponding Email: khaynes0112@gmail.com}

\begin{abstract}
We present observations of two occultations of the extrasolar planet WASP-33b using the Wide Field Camera 3 (WFC3) on the HST,  which allow us to constrain the temperature structure and composition of its dayside atmosphere. WASP-33b is the most highly irradiated hot Jupiter discovered to date, and the only exoplanet known to orbit a \ds star. We observed in spatial scan mode to decrease instrument systematic effects in the data, and removed fluctuations in the data due to stellar pulsations. The RMS for our final, binned spectrum is ~1.05 times the photon noise. We compare our final spectrum, along with previously published photometric data, to atmospheric models of WASP-33b spanning a wide range in temperature profiles and chemical compositions. We find that the data require models with an oxygen-rich chemical composition and a temperature profile that increases at high altitude. We also find that our spectrum displays an excess in the measured flux towards short wavelengths that is best explained as emission from TiO.  If confirmed by additional measurements at shorter wavelengths, this planet would become the first hot Jupiter with a temperature inversion that can be definitively attributed to the presence of TiO in its dayside atmosphere.
\end{abstract}


%-----------------------------------------------------------------------------%

\section{Introduction}
\label{I}

One of the most intriguing areas of study in the field of exoplanet characterization is the temperature structure of exoplanet atmospheres. Hot Jupiters represent an extreme end of the exoplanet distribution: they orbit very close to their host stars, which subjects them to an intense amount of stellar radiation. Also due to their proximity, they likely become tidally locked on astrophysically short timescales \citep{Guillot1996}, and are heated only on the side facing the star. This results in strong zonal winds \citep{Showman2008} that redistribute the heat, with the dynamics of this redistribution dictated by the physical and thermal structure of the planet's atmosphere.

Temperature inversions were an early prediction from atmospheric models of highly irradiated planets \citep{Hubeny2003, Fortney2008}, which demonstrated that strong absorption of incident UV/visible irradiation by high-temperature absorbers such as TiO and VO, which are commonly found in low-mass stars and brown dwarfs, could lead to thermal inversions in their observable atmospheres. Evidence for the existence of thermal inversions began with the first secondary eclipse measurements of HD209458b taken with the IRAC camera on {\it Spitzer} by \citet{Knutson2008}, who measured larger eclipse depths in spectral regions with higher opacity due to features of H$_2$O and CO (4.5 and 5.6 $\mu$m) compared with nearby bands measuring the deeper thermal continuum (3.6 and 8 $\mu$m). However, more recent analyses of  HD209458b by \citet{Diamond-Lowe2014} and \citet{Schwarz2015} do not support an inverted atmophere model; additionally, indications for the presence or absence of an inversion in other planets based on {\it Spitzer}/IRAC data appear to defy predictions based on the level of incident radiation or the overall equilibrium temperature of the atmosphere (\citealp{Machalek2008, Fressin2010a}, and others). More recent models have suggested that heavy molecules such as TiO and VO may not remain suspended in the upper atmosphere of Jupiter-mass planets \citep{Spiegel2009}, and searches for specific spectral signatures of TiO in the optical have been unsuccessful (\citealp{Sing2013}; but see also \citealp{Hoeijmakers2014} for discussion of incompleteness in the TiO line list contributing to inabilities to confirm observational detections). Recent theories have postulated several additional atmospheric processes that could play a role in the formation of inversions, such as the production of photochemical sulfur-based hazes \citep{Zahnle2009} or the inhibition of oxide formation due to a super-solar  C/O ratio \citep{Madhusudhan2012a}. Furthermore, \citet{Knutson2010a} explored the possibility that the absorbing molecular species may be destroyed by photodissociation and may, hence, be affected by the activity of the host star.

Progress on understanding the origin and conditions required for temperature inversions has been further hampered by a lack of high-quality data for most sources.  Only a single unambiguous spectrally resolved measurement of a molecular feature in the eclipse spectrum of a planet, the detection of water absorption in the WFC3 spectrum of WASP-43 b by \citet{Kreidberg2014a}, has been published to date. Eclipse measurements for most transiting exoplanets comprise only the broadband {\it Spitzer}/IRAC filters, making the conclusions largely model-dependent and subject to possible systematic offsets or uncertainties.  The inference of thermal inversions from IR photometry is based solely on our ability to determine whether there is a larger-than-expected flux from molecular bands compared with the continuum. Warm Spitzer photometry has now measured two-band eclipse depths for a large number of planets, but while these measurements can provide some indication of a potential inversion, such data cannot uniquely identify inverted atmospheres because of degeneracies between atmospheric composition and structure \citep{Madhusudhan2010, Stevenson2010, Moses2013}. In particular, \citet{Madhusudhan2010} showed that with only a few data points, this interpretation is heavily dependent on the assumed composition of the planet and the accuracy of the uncertainties ascribed to each measurement. A subsequent Bayesian retrieval analysis on a subset of well-observed planets covering a wide range of effective temperatures by \citet{Line2013} showed that the data {\bf are} inconsistent with thermal inversions for many of the planets expected to have an inversion due to the aforementioned theories for the physical origin of the phenomenon. More recently, a new analysis by \citet{Diamond-Lowe2014} and \citet{Schwarz2015} revealed that a thermal inversion is not necessary to explain the {\it Spitzer} observations of HD 209458b, previously considered the prototypical example of a planet with an inverted atmosphere. \citet{Hansen2014} has also suggested that the uncertainties on many older, single-visit {\it Spitzer} eclipse depths may be significantly higher than previously reported, resulting in data sets that are essentially consistent with featureless blackbody spectra.

It is therefore critical that we further investigate planets that provide the best chance for confirming the presence of temperature inversions, in order to better constrain the actual temperature structure of these planets and clarify the role of various stellar and planetary characteristics in defining this structure.  Here we present new secondary eclipse (or occultation) observations of WASP-33b, one of the largest and hottest planets known, using the Wide Field Camera 3 (WFC3) on HST.  WASP-33 is an A-type \ds star and its planet, WASP-33b, is one of the most highly irradiated planets discovered to date, orbiting once every 1.22 days \citep{Cameron2010, Herrero2011a}. WASP-33b is unique, being the only exoplanet yet discovered to orbit a \ds star. Multiple observations of the host star over wavelengths ranging from the visible to the infrared have shown oscillations with a range of frequencies, and amplitudes on the order of 1 mmag. Given the extreme irradiation received by WASP-33b, it is one of the most likely hot Jupiters to host a thermal inversion as TiO/VO, if present, would be expected to be in the gaseous phase throughout the observable dayside atmosphere, thereby causing a thermal inversion.

Previous occultation observations in the infrared \citep{Deming2012} concluded that WASP-33b might host a temperature inversion with a solar composition atmosphere, or a non-inverted atmosphere with enhanced carbon abundance.  The inversion scenario is advocated by \citet{DeMooij2013}, based on WASP-33b's apparent inefficient heat redistribution, which was also noted by \citet{Smith2011}. Our spectroscopic observations with WFC3 cover a wavelength range from 1.1 to 1.7 $\mu$m, which provides a valuable opportunity to confirm the presence of an inversion in WASP-33b. The WFC3 spectral range covers strong infrared molecular bands of H$_2$O and TiO, both of which are expected to be abundant in the atmosphere of WASP-33b, assuming a solar abundance composition. Assuming these molecules contribute significant opacity at the height of the thermal inversion, the presence of a thermal inversion would lead to emission features in the dayside spectrum due to both these molecules in the WFC3 range, as opposed to absorption features if no thermal inversion is present.

We describe the observations in Section 2, data reduction in Section 3, removal of stellar oscillations and analysis strategies in Section 4, and discussion of results in Section 5. 

%-----------------------------------------------------------------------------%

\section{Observations}
\label{O}

Two occultations of WASP-33 were observed on November 25, 2012 and January 14, 2013. We used WFC3's infrared G141 grism, which provides slitless spectra from 1.1\,$\mu$m to 1.7\,$\mu$m at a resolving power of 130 \citep{Dressel2012}. Each target was allocated 5 HST orbits, which was sufficient to cover a single planetary occultation while including periods of the orbit both before and after occultation.

Both sets of observations were taken using the 256 x 256 sub-array with 7 non-destructive reads per exposure, using the RAPID sampling sequence. The data were observed in spatial scan mode \citep{McCullough2012}, which increases the photon collection efficiency of the detector, and additionally has been shown to decrease systematic patterns in the data that can result from persistent levels of high flux on individual pixels. All scans were performed in the same direction. See Table \ref{obs} for details.  

\begin{deluxetable}{ccc}
\tablecaption{Observations of WASP-33}
\tablewidth{0pt}
\tablehead{\colhead{} &
                    \colhead{Visit 1} &
                    \colhead{Visit 2} }
\startdata
Time of first scan (MJD UT)   			            	   & 56256.405  & 56306.455\\
Planetary orbital phase at first scan        		   & 0.328 & 0.354 \\
Time of last scan (MJD UT)          			  	   &  56256.687 &56306.746 \\
Planetary orbital phase at last scan 		            & 0.549 &0.583\\
Number of scans                     				   & 119 & 119\\
Number of HST orbits      					   & 5 &5\\
Detector subarray size                      			   &256 & 256\\
Detector reads per scan                    			   & 7 &7\\
Duration of scan (s)                      			   & 51.7 & 51.7\\
Signal level on detector (electrons pixel $^{-1}$)  & 4.0-7.3 $\times\ 10^{4}$ &4.0-7.3 $\times\ 10^{4}$\\
\enddata
\label{obs}
\end{deluxetable}

%-----------------------------------------------------------------------------%

\section{Data Reduction}
\label{DR}
We used the series of single-exposure ``ima" images produced by the WFC3 {\bf calwf3} pipeline for our data analysis. The ``ima" files are fully reduced data products with the exception of a step to combine multiple reads. The final stage ``flt" files provided by the Space Telescope Science Institute are not appropriate for use in spatial scan mode, since the additional pipeline processing for combining multiple reads does not account for the motion of the source on the detector in spatial scan mode. We followed the methodology of \citet{Deming2013} to produce 2D spectral frames from the ``ima" files provided on MAST. We began by applying a top-hat mask in the spatial dimension of each read, the width of which is 20 pixels tall in order to fully cover the stellar PSF. Areas outside the mask were zeroed. We then subtracted subsequent reads, and then added the differenced frames to create one scanned image. We used our own strategy from \citet{Mandell2013} to search for and correct bad pixels within the combined spectral frames, and collapse the images into 1D spectra. We used the modified coefficients from \citet{Wilkins2014} to produce the wavelength and wavelength-dependent flat-field calibrations. For background correction, we subtracted a single background value from each difference pair in the WASP-33 ``ima" files before applying the top-hat mask. The background subtraction decreases the overall flux level of each light curve, thereby increasing the measured eclipse depth compared with non-background subtracted frames.  We determined the change in eclipse depth for the band-integrated light curves to be 140 and 110 ppm for Visit 1 and Visit 2, respectively; these values were constant across the spectrum.

%We identified in the direct image the nearby star noted by \citet{Adams2013}, which lies 1.9" from WASP-33. However, this nearby star has a magnitude of only $\Delta$K$_{s}$ = 5.69, which means it accounts for a flux difference of $\sim$0.5\%, well below our measured uncertainties. Due to this negligible effect on our data and the difficulty of removing such an object, especially from spatial scan data, we decided not to attempt any corrections for the additional flux during our analysis. 

We trimmed roughly 70 pixels from either end of the spectral extent, to remove the parts of the spectrum with low sensitivity. After trimming the edges of the spectrum, the spectrum covers the region between approximately 1.13 and 1.63 $\mu$m. We also identified the strong Paschen $\beta$ stellar feature at 1.28 $\mu$m, and took care to isolate it when defining our spectral bins. In an oscillating star, such spectral features may have variable line profiles, which could cause sharp changes in flux and add additional noise. 

%-----------------------------------------------------------------------------%
\section{Analysis and Results}
\label{An}

In the following sections, we describe in detail our fitting process. Briefly, we began with the band-integrated curve, meaning we summed over all wavelengths to form one photometric light curve. We used this higher sensitivity light curve to fit for wavelength-invariant parameters and determine the strongest signals due to stellar pulsations. We used the residuals of this band-integrated curve as a model for otherwise uncorrected sources of correlated noise, such as additional stellar oscillations or instrument systematics. We used binned light curves comprising (on average) 10 columns/channels in order to investigate wavelength dependent behavior, especially the change of eclipse depth with wavelength, and corrected for spectral and spatial drift of the detector with time. 

\subsection{Identifying and Fitting Trends}
\label{Sys}

\subsubsection{Stellar Oscillations}
\label{Sys}

\begin{deluxetable}{ccc}
\tablecaption{Fitted frequencies and amplitudes of WASP-33's stellar oscillations.} 
\tablewidth{0pt}
\tablehead{\colhead{Frequency} &
                    \colhead{Amplitude} &
                    \colhead{Visit} }
\startdata
19.88  $\pm$ 0.32  & 0.62 $\pm$ 0.05  & Visit 1\\ %updated September 26, 2014
29.65  $\pm$ 0.48  & 0.32 $\pm$ 0.04  & Visit 1\\
14.40  $\pm$ 0.56  & 0.65 $\pm$ 0.16  & Visit 2\\
22.16  $\pm$ 0.57  & 0.50 $\pm$ 0.07  & Visit 2\\
\enddata
\label{osc_modes}
\end{deluxetable}

WASP-33 is known to be an oscillating \ds star whose pulsation frequencies have been measured over multiple campaigns \citep{Herrero2011a,Smith2011,Deming2012,Sada2014,DeMooij2013,Kovacs2013,Essen2013} across a wide range of wavelengths, and many different pulsation frequencies have been determined by these studies. However, since multiple oscillation modes are to be expected, and that the strength of these modes will vary with wavelength, observations taken across a range of spectral bands and at various times should not be expected to have perfect agreement, nor can we expect exact comparisons across data sets. The incomplete temporal sampling caused by HST orbits complicates characterization of the oscillation modes in our data, and so we explored different avenues for constraining the detectable pulsation frequencies and removing the stellar oscillations. 

We used sine functions to model the stellar pulsations. While a non-parametric approach such as Gaussian Processes might allow more accurate modeling of stellar pulsations, which can be quasi-periodic, previous observing campaigns using parametric aproaches did not suffer from the incomplete temporal sampling in our data, and so by following their example we were able to make comparisons between our best-fit frequencies and more complete datasets. 

We divided frequency space into regions based on the frequencies identified by previous observing campaigns and allowed our MCMC models to fit, iteratively or simultaneously, between 1 and 3 sine curves restricted to those regions of frequency space. We used Bayesian information criterion (BIC;  \citealp{Schwarz1978, Liddle2004}) to determine the best combination of sine terms. While previous measurements have shown that the phase of stellar oscillations can change slightly with wavelength \citep{Conidis2010}, the potential amplitude of these changes would be very small across the WFC3 wavelength range, and we did not attempt to fit for any phase change with wavelength. We find that two sine curves achieve the best results without overfitting the data, and that the frequencies and amplitudes identified are robust whether we fit the sine curves simultaneously or in sequence. The results are shown in Table \ref{osc_modes}, and our best-fit frequencies agree approximately with previously determined values. 

Red noise remains in the residuals after removal of the two sine curves representing the stellar oscillation modes, indicating that we are unable to fully characterize either the stellar oscillations or underlying instrument systematics. The instrument systematics are weak in spatial scan mode, but still present \citep{Deming2013}. However, as we describe in later sections, the remaining red noise in the band-integrated light curve will not affect our relative wavelength-dependent eclipse depths since we subtract a scaled version of the residual noise from each bin. Additionally, because the first orbit of WFC3 observations tends to be more noisy than subsequent orbits \citep{Mandell2013, Deming2012}, we do not include this orbit for the band-integrated fitting process (including fitting for the stellar oscillations), though we do incorporate it later in our wavelength-dependent relative analysis. 

\subsubsection{Nonlinearity Correction}
\label{Sys}

WASP-33 is an early-type star, and the flux incident on the detector at the shortest wavelength of the grism response is almost a factor of two larger than the flux at the longest wavelength.  Since we want to optimize the photon-limited SNR at even the longest wavelength, we exposed the spectra to the highest possible fluence levels, reaching $7 \times 10^4$ electrons per pixel at the short wavelength end.  Since this is comparable to the full well value of the detector, we applied a correction for detector nonlinearity, as described in Sec. 6.5.1 of the WFC3 Data Handbook 2.1.  We used coefficients valid for quadrant one of the detector, extracted from the calibration files at STScI.  Our calculated correction increases the eclipse depths by 25 parts per million at our shortest wavelength, decreasing to about 6 parts per million at the longest wavelength we analyze.  These corrections do not affect our scientific conclusions, which would be virtually identical if we had omitted the non-linearity correction.

\subsubsection{Spectral Shifts With Time}
\label{Sys}

In order to correct for possible variations in channel flux due to shifts across the peak of spectral features with time, we calculated the shift in the horizontal/spectral direction referenced to a template exposure from our data. We examined two main strategies for measuring the magnitude of the shift with time. Our initial method was to measure the spectrum at the steeply sloped edges of the grism response. Due to the steep slope, sensitivity due to spectral shifting is greatest at these wavelengths; however, these wavelengths mostly do not overlap with those used in our final analysis, which are located across the central, flatter region of the grism response. If the shift of spectrum is identical at all wavelengths, this should not impact the final results; however, as we describe below, this was not the case. The alternative strategy is to use only those wavelengths also used for the subsequent light curve analysis. We describe our analysis of both strategies in greater detail, along with their results, below. 

In our initial analysis procedure, which was adapted from the method used in \citet{Mandell2013}, we fit the slope of the pixels at the short-wavelength and long-wavelength edges of the spectrum. At these wavelengths the sensitivity curve causes the shape of the spectrum to change most dramatically, resulting in a high-precision measurement of the spectral shift in each exposure. The slopes measured at both edges of the spectrum were averaged to further decrease the effective uncertainty of the measurement for each exposure. We used the zeroth-order coefficient of this fit to determine the shift of each spectrum relative to the first exposure in the time series. 

We compared this strategy to that employed by \citet{Deming2013}, which used instead the central region of the spectrum to measure the shifts. This region is flat compared to the edges, but does contain modulation in the spectral response. In this method, we created a template spectrum comprised of an average of the exposures in the time series immediately preceding and following eclipse.  We interpolated the template spectrum onto a wavelength grid shifted in either direction up to a pixel and a half, stepping in 0.001 pixel increments, and saved each shifted spectrum. For each exposure, we compared the observed spectrum with each shifted template spectrum. We also allowed the template spectrum a linear stretch in intensity.  We calculated the rms for each comparison; the shift corresponding to the lowest rms is saved as our best-fit spectral shift. 

\begin{figure}[htbp]
\centering
{
\includegraphics[width=175mm]{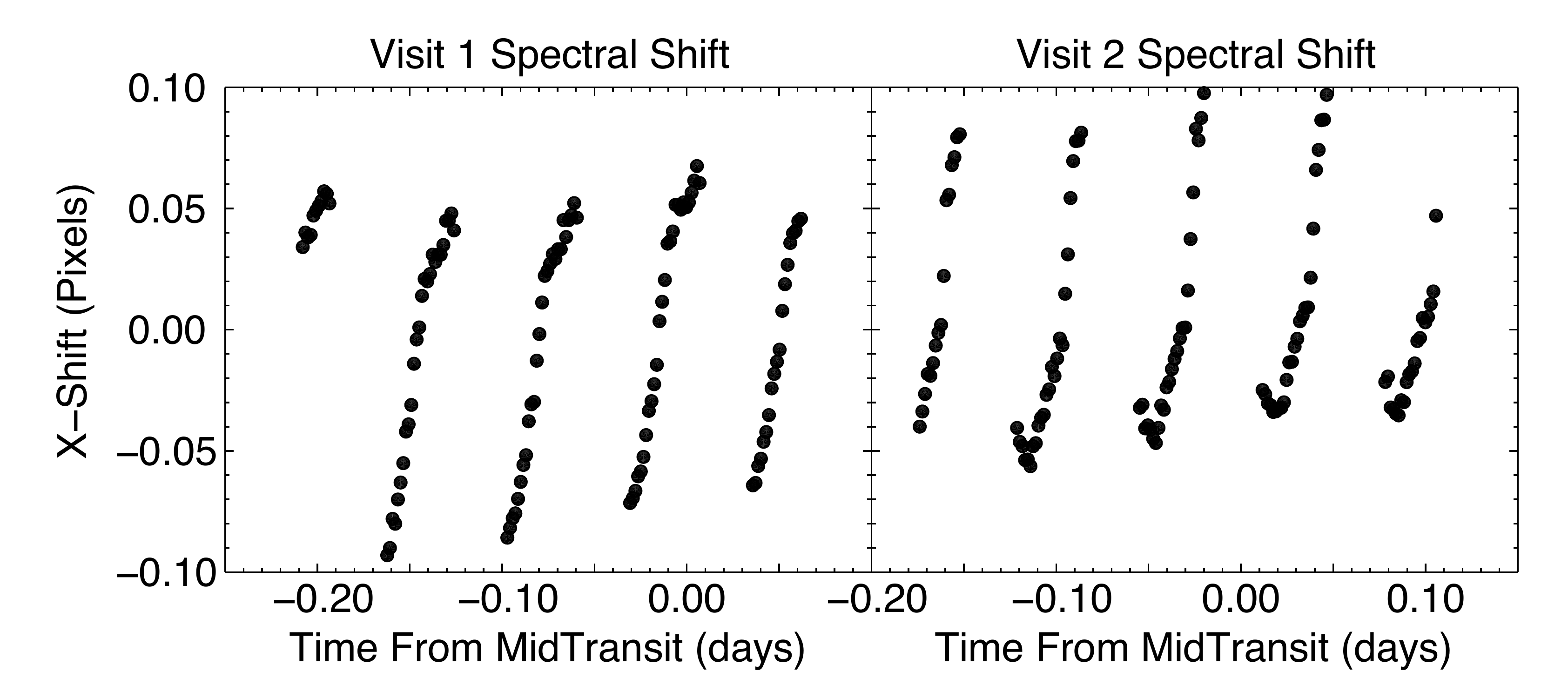}
\caption{Shifts in the spectral direction of the detector for both visits across the eclipse duration. These shifts have been measured using the entire central region of the WFC3 wavelength coverage. }
 \label{fig:shifts}
}
\end{figure}

We compared the two spectral shift measurement strategies, finding that if the edges of the spectrum are included for the latter method, then the resulting shifts match the ``edge only" measurements of our method very closely. If instead only the central region of the spectrum is used to determine the shifts, the measured shifts result in significant change in the final eclipse spectrum. This result indicates that the shape or placement of the grism sensitivity function changes as a function of time or placement on the detector, possibly due to changes in the optical path as part of the thermal breathing modes of the telescope. The horizontal shifts produced from the entire central region for the two visits are plotted in Figure \ref{fig:shifts} for reference. The use of the ``center-only'' spectral shifts resulted in a lower residual rms for the resulting light curves, indicating that the fit was improved by using shifts derived from the same portion of the spectrum as the light curves themselves.

\begin{figure}[htbp]
\centering
{
\includegraphics[width=175mm]{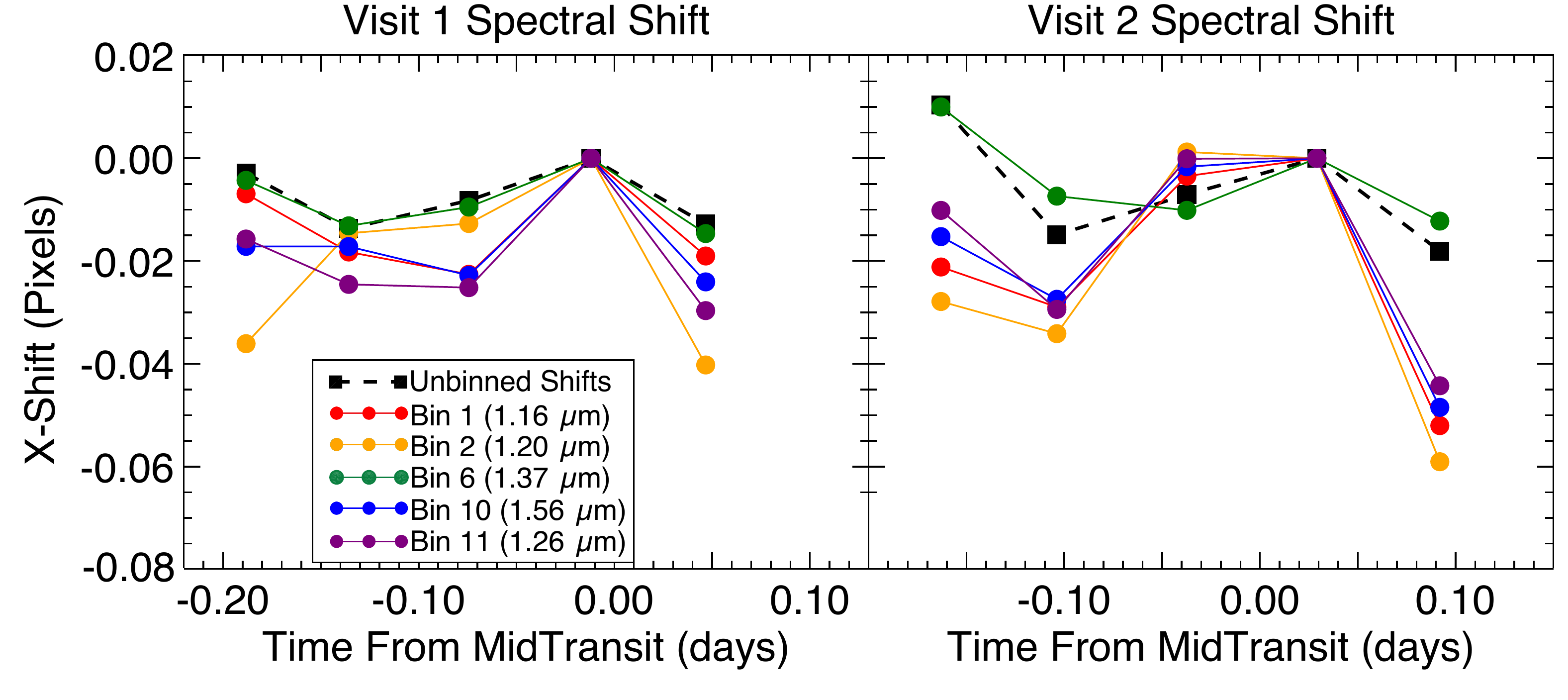}
\caption{Shifts in the spectral direction of the detector, binned in time to show one data point per HST orbit. The shifts measured from the entire central wavelength region ("unbinned shifts") are shown as black boxes. Refer to Figure \ref{fig:shifts} for the same data, unbinned in time. A selection of "binned shifts" are shown as colored circles. In this case, the spatial shifts on the detector are measured for individual bins in wavelength, comprising about 10 channels/columns. While the bin from the most central region of the spectrum tracks closely with the unbinned shifts, the bins from closer to the edges of the spectrum show trends that differ both from the central region, and from visit to visit, suggesting that the shape or placement of the grism sensitivity function is changing over the duration of a single visit. }
 \label{fig:shifts_multi1}
}
\end{figure}

For consistency, we further extended this analysis by determining the specific spectral shift for the specific portion of the spectrum associated with each binned light curve, and using that set of shifts in the systematic decorrelation procedure. This has the advantage of using the set of shifts that best describe the region of the spectrum used by each bin, since the stellar intensity can vary greatly across the WFC3 grism response. However, the modest wavelength range of the WFC3 grism and the minimal temperature variations expected due to stellar oscillations ensure that our results show no spectral response to these pulsating modes. As seen in Figure \ref{fig:shifts_multi1}, the overall trend of the shifts with time does change with wavelength, at least on a visit-long level; finer analysis is not possible due to the scatter in the measurements. We found that using the bin-specific spectral shifts resulted in final spectra for both visits that were consistent within uncertainties; on the contrary, using a single shift value led to larger deviations between the visits, especially at short wavelengths.  Given this, we advocate careful inspection of the shift of the spectrum on the detector with time, and for this work, we use these binned shifts for our final analysis. 

In all cases, the shifts take the form of a repeating, inter-orbit pattern, as well as a visit-long slope. We removed a visit-long linear trend from the xshifts, since a visit-long trend in flux is also seen in previous WFC3 data sets and we choose to instead fit for this slope as an independent parameter in our light curve fit. Alternatively, we also examined the shift correction strategy from \citet{Deming2013}, in which we interpolate each exposure's spectrum onto a shifted wavelength grid according to its best-measured shift value (as determined by the rms) and use these shifted spectra for light curve fitting. In this case, we do not use the scaled xshifts as a parameter in our MCMC fitting; instead, we shift the spectra themselves. We find that the final derived spectrum is minimally affected by the method of correction (interpolation vs xshift scaling), so long as the same wavelength ranges (center, center + edges, or binned central region) are used to measure the shifts for both methods.

\subsection{Band-Integrated Eclipse Curve Fitting}
\label{WLCF}

We fit a band-integrated eclipse curve first, in order to determine best-fit values for those parameters that are not wavelength dependent (as in the case for the stellar oscillations), and in order to use the residuals of this band-integrated fit as a component in our analysis of spectral channels or bins (see \citet{Mandell2013} for further explanation and details of our fitting process). For both band-integrated and spectral channels/bins, we use a Markov Chain Monte Carlo (MCMC) routine to determine our best-fit parameters. We locked all orbital parameters, leaving open for fitting only the eclipse depth, slope, and sine terms. Orbital parameters are listed in Table \ref{p2litvals}.

\begin{deluxetable}{cc}
\tablecaption{Orbital and Stellar Parameters for WASP-33 \tablenotemark{a}}
\tablewidth{0pt}
\tablehead{\colhead{Parameter} &
                   \colhead{Value} }
\startdata
Period (days)					& 1.2198709   \\
$i$ ($^{\circ}$)	 				& 86.2 $\pm$ 0.2	\\	
$R_{p}/R^{*}$  					& 0.1143 $\pm$ 0.0002  \\	
$a/R^{*}$ 						& 3.69 $\pm$ 0.01  \\	
Semi-major axis (AU) 			& 0.0259 $^{+0.0002}_{-0.0005}$  \\	
e                						& 0.00 $\pm$ 0.00   \\
\hline
$M_{*}$ (M$_{\odot}$)	 		& 1.561$^{+0.045}_{-0.079}$\\	
Spectral type	 				& A5 	 \\	
H band Magnitude			         & 7.5    \\
\lbrack Fe/H\rbrack 				& 0.1 $\pm$ 0.2 \tablenotemark{c}\\	
\enddata
\label{p2litvals}
\tablenotetext{a}{Values from \citet{Kovacs2013} except where otherwise noted.}
\tablenotetext{c}{From \citet{Cameron2010}.}
\end{deluxetable}

\begin{figure}[htbp]
\centering
{\includegraphics[width=140mm]{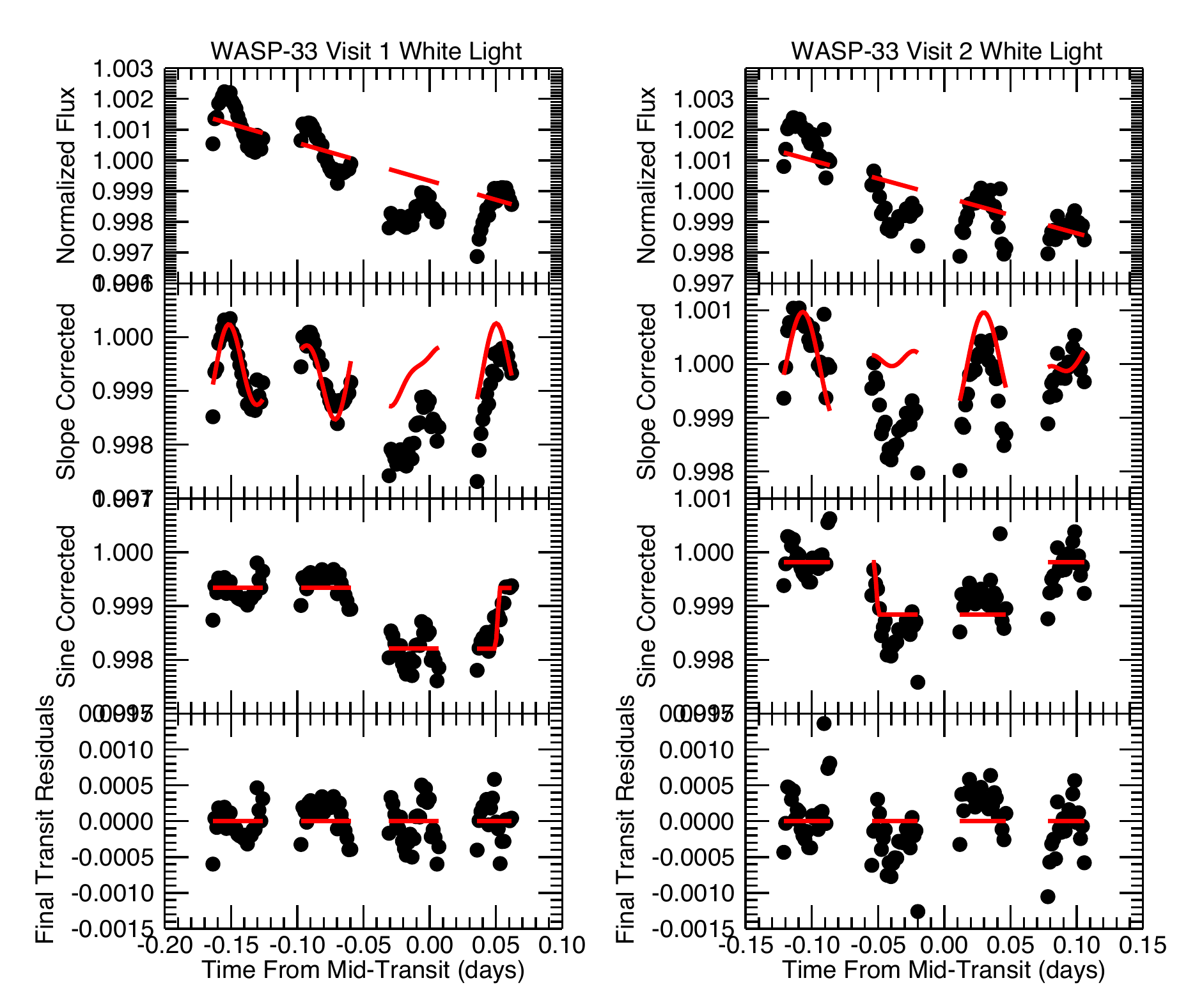}
\caption{Band-integrated light curves for both visits in black with various model fits in red. From top to bottom, the plots show a) the normalized data with a model for the visit-long slope effect; b) the slope corrected data with a model for the stellar oscillations; c) the slope and oscillation corrected data with a model for the eclipse; d) the residuals for the full fit. Parameters are fit concurrently in MCMC, and are shown here in stages for clarification on the relative contributions of each parameter. 
}
 \label{fig:whl_stages}
}
\end{figure}

Our band-integrated time series is shown in various stages in Figure \ref{fig:whl_stages}. While in practice we fit all parameters simultaneously, we show here the effect of removing systematics one at a time, and overplotting models for the slope, stellar oscillations, and finally the eclipse model itself on iterative versions of the residuals. Due to the offset in the phase of the stellar oscillations from one visit to the next, we fit each visit separately. We find agreement for the two visits' eclipse depths at the $\approx 1.5 \sigma$ level, suggesting the impact of the remaining red noise does not affect the determined eclipse depths significantly. We take for our final, best fit eclipse depth the weighted mean of both visits; these values are listed in Table \ref{p2whl}. 

\begin{deluxetable}{cc}
\tablecaption{Band Integrated Results}
\tablewidth{0pt}
\tablehead{\colhead{Visit} &
                   \colhead{Eclipse Depth (\%)} }
\startdata
Visit 1						& 0.129 $\pm$ 0.009 \\ %updated January 11, 2015, now includes Drake's background correction
Visit 2		 				& 0.110 $\pm$ 0.010  \\	
Combined	 				& 0.119 $\pm$ 0.006  \\		
\enddata
\label{p2whl}
\end{deluxetable}

\subsection{Fitting the Spectrally Binned Light Curves}
\label{BLCF}

For our wavelength-dependent analysis, we continued to fit each visit separately. However, for each visit we created a modified set of residuals from the band-integrated curve fitting step by using one joint eclipse depth derived from the weighted mean of both visits, rather than the individual best-fit eclipse depth from each visit. This provides a common band-integrated eclipse depth and standardizes the offset for each visit. Following the methodology of \citet{Mandell2013}, for each visit we incorporated into our model for the eclipse of each spectral bin the residuals from the band-integrated light curve as well as a systematic trend based on our measurements of the horizontal shift of the spectrum on the detector over time. Both of these components were allowed a scaling factor which we left open as an additional fitting parameter.

For the light curves of each channel or bin of channels we followed the band-integrated methods for fitting using MCMC. We locked the same orbital parameters, and used a BIC to determine whether the sine amplitudes, second order polynomial coefficient, residuals scaling, and spectral shift scaling terms should be varied in our final analysis. The results of open model parameters for individual bins are shown in Tables \ref{BIC_res1} and \ref{BIC_res2}.  

\begin{deluxetable}{c|c|c|c|c|c|c|c}
\tabletypesize{\scriptsize}
\tablecaption{Visit 1: Open parameters for each bin as determined by BIC. For each binned light curve, we test the significance of additional parameters with a BIC. Parameters determined to significantly improve the fit are marked with a checkmark. We also tested  a residuals scaling coefficient, but found that it was not significant for any bin, and so we do not include the null result in our table.}
\tablewidth{0pt}
\tablehead{\colhead{Wavelength } &
                   \colhead{Eclipse } &
                   \colhead{Out of } &
                   \colhead{Linear} &
                   \colhead{Sine  } &
                   \colhead{XShifts  } &
                   \colhead{2nd Order}\\
                   \colhead{($\mu$m)} &
                   \colhead{ Depth} &
                   \colhead{Eclipse Flux} &
                   \colhead{Slope} &
                   \colhead{Amplitudes } &
                   \colhead{Scaling } &
                   \colhead{Coeffecient}
                   }
\startdata
1.16 & \checkmark & \checkmark &\checkmark & &\checkmark & \checkmark \\
1.20 & \checkmark & \checkmark &\checkmark &\checkmark &\checkmark & \\
1.24 & \checkmark & \checkmark &\checkmark & &\checkmark &\checkmark \\
1.28 & \checkmark & \checkmark &\checkmark & &\checkmark & \\
1.32 & \checkmark & \checkmark &\checkmark & &\checkmark &  \\
1.37 & \checkmark & \checkmark &\checkmark & &\checkmark &  \\
1.41 & \checkmark & \checkmark &\checkmark & &\checkmark &\checkmark \\
1.46 & \checkmark & \checkmark &\checkmark & &\checkmark &\checkmark \\
1.51 & \checkmark & \checkmark &\checkmark & &\checkmark & \checkmark  \\
1.56 & \checkmark & \checkmark &\checkmark & &\checkmark &\checkmark\\
1.61 & \checkmark & \checkmark &\checkmark & &\checkmark &\checkmark \\
\enddata
\label{BIC_res1}
\end{deluxetable}

\begin{deluxetable}{c|c|c|c|c|c|c|c}
\tabletypesize{\scriptsize}
\tablecaption{Visit 2: Open parameters for each bin as determined by BIC. For each binned light curve, we test the significance of additional parameters with a BIC. Parameters determined to significantly improve the fit are marked with a checkmark. We also tested  a residuals scaling coefficient and a second order polynomial term, but found that they were not significant for any bins, and so we do not include the null results in our table.}
\tablewidth{0pt}
\tablehead{\colhead{Wavelength } &
                   \colhead{Eclipse } &
                   \colhead{Out of } &
                   \colhead{Linear} &
                   \colhead{Sine  } &
                   \colhead{XShifts  }\\
                   \colhead{($\mu$m)} &
                   \colhead{ Depth} &
                   \colhead{Eclipse Flux} &
                   \colhead{Slope} &
                   \colhead{Amplitudes } &
                   \colhead{Scaling } 
                   }
\startdata
1.16 & \checkmark & \checkmark& \checkmark& &\checkmark  \\
1.20 & \checkmark& \checkmark&\checkmark &  & \checkmark  \\
1.24 & \checkmark  & \checkmark&\checkmark & & \checkmark   \\
1.28 & \checkmark  &\checkmark &\checkmark &\checkmark &\checkmark   \\
1.32 & \checkmark &\checkmark &\checkmark  & & \checkmark   \\
1.37 & \checkmark &\checkmark &\checkmark & & \checkmark  \\
1.41 & \checkmark &\checkmark &\checkmark & \checkmark & \checkmark   \\
1.46 & \checkmark &\checkmark &\checkmark  & & \checkmark   \\
1.51 & \checkmark &\checkmark &\checkmark & &\checkmark    \\
1.56 & \checkmark&\checkmark &\checkmark & & \checkmark  \\
1.61 & \checkmark&\checkmark &\checkmark & & \checkmark \\
\enddata
\label{BIC_res2}
\end{deluxetable}

In general we find that the same terms make significant contributions for all the bins in a single visit. The light curves are typically fit best by a model including a linear slope term, unscaled band-integrated residuals and sine amplitudes, and a scaled version of the spectral shifts. A fit for a second-order polynomial term for the visit-long trend (as suggested by \citealp{Stevenson2014}) passed the BIC for most bins in Visit 1, but did not pass for any bins in Visit 2.  In all cases we chose the open parameters based on the BIC, on a bin-by-bin basis. 

\begin{figure}[htbp]
\centering
{
\includegraphics[width=135mm]{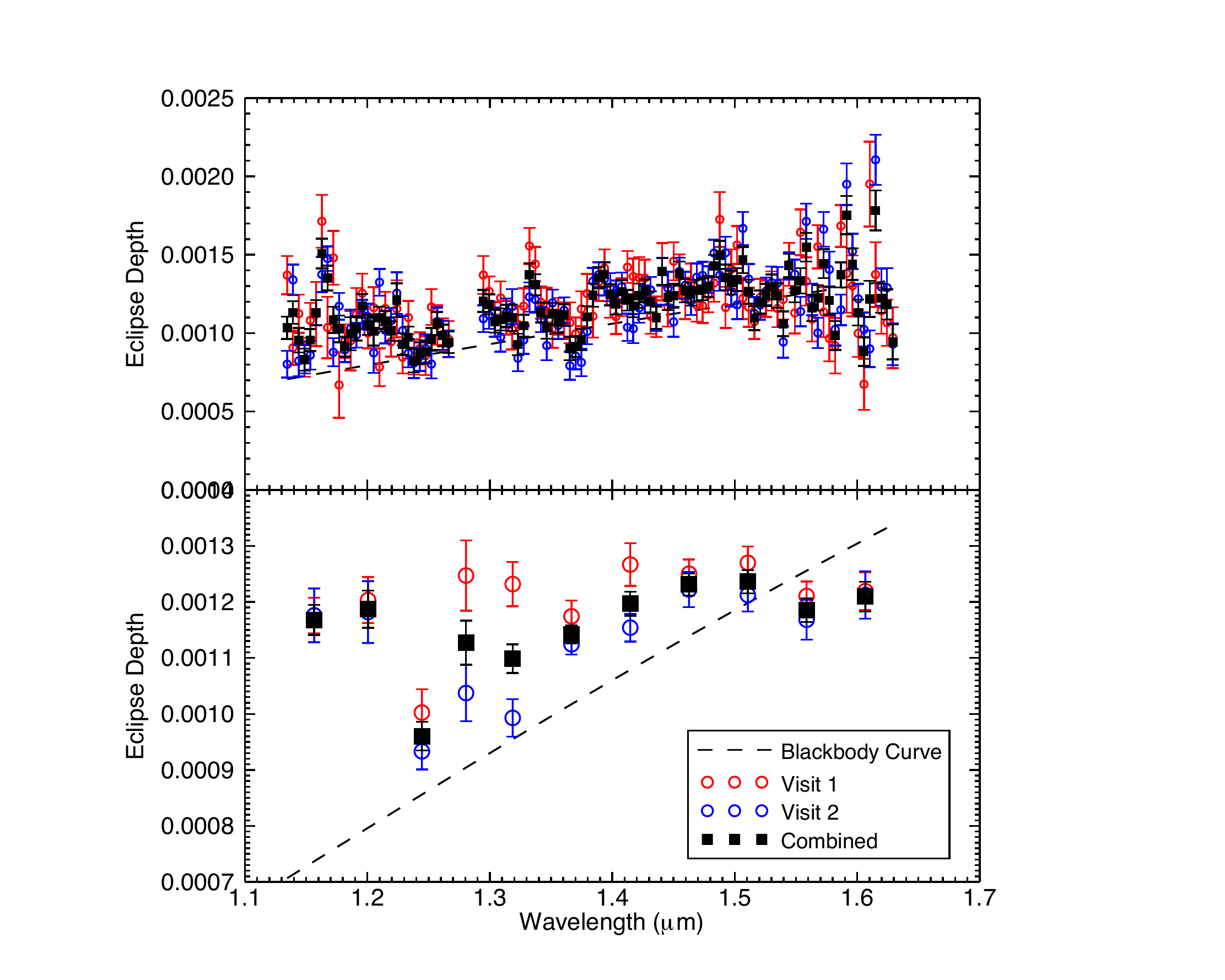}
\caption{Top: Unbinned spectrum. Here, every detector column was assigned a wavelength, a light curve was extracted, and the eclipse depth measured. Bottom: Binned spectrum. In this case, multiple detector columns (ten columns, on average) were combined to make one light curve, and the eclipse depth measured. Points are plotted at the mean wavelength for each bin. For both plots, Visit 1 is in red, Visit 2 in blue, and the combined visits are in black. Combined visits use a weighted mean. The largest discrepancies  between visits are seen near the stellar hydrogen line at 1.28 $\mu$m. Blackbody curve is shown as a dashed line, with the best-fit planetary temperature of 2950 K. }
 \label{fig:combo_spec}
}
\end{figure} 

Finally, we used an uncertainty-weighted mean to combine both visits in our final stage of analysis, and these results are presented in Figure \ref{fig:combo_spec} and in Tables \ref{ch_data} (for the channels) and \ref{bin_data} (for the bins).  The results from both visits agree to within the 1-$\sigma$ uncertainties for almost all of the spectral bins, with the only major discrepancies appearing for two bins near the stellar hydrogen line at 1.28 $\mu$m.  This difference between visits for the hydrogen feature is most likely due to variability in the depth of the stellar hydrogen line, which can oscillate differently than the overall stellar continuum; we therefore consider this offset to be uncorrectable without simultaneous measurements of other hydrogen spectral features. 

\clearpage
\begin{deluxetable}{cc|cc|cc}
\tabletypesize{\scriptsize}
\tablecaption{Spectral Results for Channel Data} %updated January 11, 2015
\tablewidth{0pt}
\tablehead{\colhead{Wavelength } &
                   \colhead{Eclipse Depth} &
                   \colhead{Wavelength } &
                   \colhead{Eclipse Depth } &
                   \colhead{Wavelength } &
                   \colhead{Eclipse Depth }  \\
                   \colhead{ ($\mu$m)} &
                   \colhead{ (\%)} &
                   \colhead{ ($\mu$m)} &
                   \colhead{ (\%)} &
                   \colhead{ ($\mu$m)} &
                   \colhead{ (\%)}
                    }
\startdata
1.135 & 0.105 $\pm$ 0.00714 &  1.304 & 0.110 $\pm$ 0.0071  &   1.474 & 0.130 $\pm$ 0.00695 \\
1.139 & 0.115  $\pm$ 0.00725 &  1.309 & 0.111 $\pm$ 0.0069  &   1.478 & 0.132 $\pm$ 0.00678 \\
1.144 & 0.0973 $\pm$ 0.00779 &  1.314 & 0.113  $\pm$ 0.0070  &   1.483 & 0.145 $\pm$ 0.00699 \\
1.149 & 0.0844 $\pm$ 0.00706 &  1.318 & 0.112  $\pm$ 0.0084  &   1.488 & 0.152 $\pm$ 0.00903 \\
1.153 & 0.0973 $\pm$ 0.00709 &  1.323 & 0.0946 $\pm$ 0.0068  &   1.493 & 0.138 $\pm$ 0.00817 \\
1.158 & 0.115  $\pm$ 0.00815 &  1.328 & 0.107 $\pm$ 0.0070  &   1.497 & 0.134 $\pm$ 0.00767 \\
1.163 & 0.154  $\pm$ 0.00953 &  1.332 & 0.140  $\pm$ 0.00732 &   1.502 & 0.136 $\pm$ 0.00735 \\
1.168 & 0.138  $\pm$ 0.00791 &  1.337 & 0.134  $\pm$ 0.00667 &   1.507 & 0.149 $\pm$ 0.00835 \\
1.172 & 0.111 $\pm$ 0.00833 &  1.342 & 0.115  $\pm$ 0.00632 &   1.511 & 0.127 $\pm$ 0.00764 \\
1.177 & 0.105 $\pm$ 0.00858 &  1.346 & 0.106 $\pm$ 0.0071  &   1.516 & 0.111 $\pm$ 0.00764 \\
1.182 & 0.0932 $\pm$ 0.00641 &  1.351 & 0.1014  $\pm$ 0.00698 &   1.521 & 0.121 $\pm$ 0.00736 \\
1.186 & 0.102 $\pm$ 0.0106  &  1.356 & 0.109 $\pm$ 0.0070  &   1.525 & 0.125 $\pm$ 0.00708 \\
1.191 & 0.105 $\pm$ 0.00761 &  1.361 & 0.1014  $\pm$ 0.0068  &   1.530 & 0.130 $\pm$ 0.00805 \\
1.196 & 0.119  $\pm$ 0.00758 &  1.365 & 0.0918 $\pm$ 0.0075  &   1.535 & 0.125 $\pm$ 0.00783 \\
1.200 & 0.108 $\pm$ 0.00637 &  1.370 & 0.0930 $\pm$ 0.0066  &   1.540 & 0.107 $\pm$ 0.00791 \\
1.205 & 0.103 $\pm$ 0.00909 &  1.375 & 0.0975 $\pm$ 0.0069  &   1.544 & 0.146 $\pm$ 0.00747 \\
1.210 & 0.112  $\pm$ 0.00705 &  1.379 & 0.112  $\pm$ 0.0069  &   1.549 & 0.128 $\pm$ 0.00763 \\
1.215 & 0.110 $\pm$ 0.00663 &  1.384 & 0.126  $\pm$ 0.00740 &   1.554 & 0.134 $\pm$ 0.00809 \\
1.219 & 0.103 $\pm$ 0.00623 &  1.389 & 0.137  $\pm$ 0.00683 &   1.558 & 0.156 $\pm$ 0.00917 \\
1.224 & 0.124  $\pm$ 0.0108  &  1.394 & 0.139  $\pm$ 0.00786 &   1.563 & 0.117 $\pm$ 0.00839 \\
1.229 & 0.0950 $\pm$ 0.00697 &  1.398 & 0.127  $\pm$ 0.00673 &   1.568 & 0.124 $\pm$ 0.00805 \\
1.233 & 0.0994 $\pm$ 0.00693 &  1.403 & 0.121  $\pm$ 0.00669 &   1.573 & 0.145 $\pm$ 0.00924 \\
1.238 & 0.0841 $\pm$ 0.00736 &  1.408 & 0.129  $\pm$ 0.00730 &   1.577 & 0.122 $\pm$ 0.00905 \\
1.243 & 0.0889 $\pm$ 0.00683 &  1.412 & 0.123  $\pm$ 0.00663 &   1.582 & 0.0993 $\pm$ 0.00919 \\
1.248 & 0.0898 $\pm$ 0.00736 &  1.417 & 0.119  $\pm$ 0.00722 &   1.587 & 0.138 $\pm$ 0.00796 \\
1.252 & 0.0980 $\pm$ 0.00718 &  1.422 & 0.126  $\pm$ 0.00779 &   1.591 & 0.176 $\pm$ 0.0123 \\
1.257 & 0.108 $\pm$ 0.00736 &  1.427 & 0.129  $\pm$ 0.00716 &   1.596 & 0.145 $\pm$ 0.0101 \\
1.262 & 0.100 $\pm$ 0.00687 &  1.431 & 0.122  $\pm$ 0.00670 &   1.601 & 0.114 $\pm$ 0.00829 \\
1.266 & 0.0960 $\pm$ 0.00689 &  1.436 & 0.112 $\pm$ 0.0071  &   1.606 & 0.0889 $\pm$ 0.00916 \\
1.271 & 0.0582 $\pm$ 0.0101  &  1.441 & 0.141  $\pm$ 0.00849 &   1.610 & 0.123 $\pm$ 0.0115 \\
1.276 & 0.0305 $\pm$ 0.00960 &  1.445 & 0.125  $\pm$ 0.00811 &   1.615 & 0.179 $\pm$ 0.0128 \\
1.281 & 0.245  $\pm$ 0.0252  &  1.450 & 0.126  $\pm$ 0.00760 &   1.620 & 0.123 $\pm$ 0.00930 \\
1.285 & 0.179  $\pm$ 0.0184  &  1.455 & 0.140  $\pm$ 0.00764 &   1.624 & 0.119 $\pm$ 0.00955 \\
1.290 & 0.121  $\pm$ 0.00934 &  1.460 & 0.129  $\pm$ 0.00742 &   1.629 & 0.0951 $\pm$ 0.0112 \\
1.295 & 0.123  $\pm$ 0.00708 &  1.464 & 0.127  $\pm$ 0.00738 &   	 & \\
1.299 & 0.120  $\pm$ 0.00647 &  1.469 & 0.129  $\pm$ 0.00681 &  		  & \\
\enddata
\label{ch_data}
\end{deluxetable}
\clearpage

\begin{deluxetable}{cc}
\tablecaption{Spectral Results for Binned Data} %updated January 11, 2015
\tablewidth{0pt}
\tablehead{\colhead{Wavelength ($\mu$m)} &
                   \colhead{Eclipse Depth (\%)} }
\startdata
1.155 & 0.119   $\pm$ 0.00270 \\
1.199 & 0.121   $\pm$ 0.00334 \\
1.243 & 0.098 $\pm$ 0.00257 \\
1.279 & 0.115   $\pm$ 0.00393 \\
1.318 & 0.112 $\pm$ 0.00256 \\
1.366 & 0.116   $\pm$ 0.00157 \\
1.414 & 0.122   $\pm$ 0.00214 \\
1.462 & 0.125   $\pm$ 0.00198 \\
1.510 & 0.125   $\pm$ 0.00207 \\
1.558 & 0.120   $\pm$ 0.00212 \\
1.606 & 0.122   $\pm$ 0.00266 \\
\enddata
\label{bin_data}
\end{deluxetable}

\subsection{Error Analysis}
\label{EA}

\begin{deluxetable}{lcc}
\tabletypesize{\scriptsize}
\tablecaption{Error Analysis. We compare photon noise, rms statistics, and both predicted and measured uncertainties for each source. We show each of these statistics for the band-integrated time series, binned data, and spectral channels. For the binned data and spectral channels, the mean uncertainty is shown for each row. The predicted uncertainty calculation assumes monotonic temporal spacing, which is not the case for WFC3 observations, and is therefore likely to underpredict the true uncertainty. The measured uncertainty is drawn from the MCMC posterior distributions. The difference between the predicted uncertainty and the uncertainty from MCMC therefore mostly reflects the incomplete coverage over the light curve, while the difference between the rms and the photon noise reflects the impact of additional noise beyond the photon noise.}
\tablewidth{0pt}
\tablehead{\colhead{Parameters} &
                    \colhead{Visit 1} &
                    \colhead{Visit 2}}    
\startdata
Data points during eclipse & 38  & 47 \\
Data points out of eclipse & 81 & 72 \\
\hline
 \multicolumn{2}{c}{Channels} \\ \hline
Photon noise (ppm) 									& 419. & 420. \\
RMS of residuals (ppm) 								& 440. & 437. \\
Predicted\tablenotemark{1} $\sigma_{ed}$ (ppm)  			& 79.0 & 78.0 \\
$\sigma_{ed}$ from MCMC+RP Data (ppm)				& 203. & 121. \\
RMS/photon noise 									& 1.05 & 1.04 \\
MCMC Data/Pred. 									& 2.57 & 1.55 \\
\hline
 \multicolumn{2}{c}{0.042-Micron Bins (11 Total)} \\ \hline
Photon noise (ppm) 									& 181. & 162. \\
RMS of residuals (ppm) 								& 191. & 187. \\
Predicted\tablenotemark{1} $\sigma_{ed}$ (ppm)  			& 34.0 & 30.1 \\
$\sigma_{ed}$ from MCMC+RP Data (ppm)				& 76.2 & 52.6 \\
RMS/photon noise 									& 1.07 & 1.17 \\
MCMC Data/Pred.									& 2.33 & 1.79 \\
\hline
 \multicolumn{2}{c}{Band Integrated Time Series} \\ \hline
Photon noise (ppm) 									& 47.3 & 46.6 \\
RMS of residuals (ppm) 								& 272. & 331. \\
Predicted\tablenotemark{1} $\sigma_{ed}$ (ppm)  			& 9.50 & 9.58 \\
$\sigma_{ed}$ from MCMC+RP Data (ppm)				& 112. & 90.5 \\
RMS/photon noise 									& 5.75 & 7.10 \\
MCMC Data/Pred. 									& 11.8 & 9.45 \\
\enddata
\tablenotetext{1}{Calculated from the photon noise and the number of points during eclipse}
\label{error}
\end{deluxetable}

Our measured uncertainties were initially drawn from the MCMC posterior probability distributions.  In order to estimate the impact of our red noise, we used a modified version of the residuals permutation method \citep{Gillon2007}, which involves shifting (permutating) the time series of the residual noise via the ``prayer-bead" method. In addition to shifting the residual noise series left over from subtracting the light-curve model, we also inverted our residuals (multiplying by -1) and reversed both the inverted and non-inverted residuals in time to produce four different sets of residual noise. This yields 4 $\times$ N permutations, where N is the number of exposures. We deemed this extra step useful because of the otherwise limited number of possible permutations, which did not yield clear results from a traditional residuals permutation analysis. For each channel or bin of channels, we use whichever is higher, the uncertainty from MCMC or residuals permutation. For Visit 1, we find that uncertainties from residuals permutation are on average 1.47 times higher than uncertainties from MCMC, while for Visit 2 uncertainties from residuals permutation are 1.20 times higher.

We compared the photon noise to the rms of our white light, channel, and binned data.We present our results in Table \ref{error}. In general we find that a substantial amount of red noise remains in our band integrated light curves after removal of our best fit models. However, the temporal morphology of the red noise does not change with wavelength, and by subtracting our band-integrated residuals from each bin we are able to closely approach the photon noise limit for our channels and bins. For our channels and bins, we find an rms $\sim$1.05 times the photon noise. 

We also compare the measured MCMC + residuals permutation uncertainty to a predicted eclipse depth uncertainty. This prediction is based on the photon noise (or rms) and the number of exposures in eclipse versus out of eclipse. The predicted uncertainty calculation assumes monotonic temporal spacing, which is not the case for WFC3 observations. We find that our measured eclipse depth uncertainties for the channels and bins are between $\sim$1.5-2.5 times the predicted uncertainty, and we ascribe this to the contributions from the uneven sampling of HST orbits and the residual red noise in the light curves. In previous studies we found that the lack of complete and evenly spaced temporal coverage during the eclipse is likely the cause of this failure to meet the predicted uncertainty, even for photon-limited results. In this same comparison of measured versus predicted uncertainties, we note that Visit 1 is further from the predicted uncertainty than Visit 2. Visit 1 has substantially fewer points post-eclipse than Visit 2, which  can be detrimental to the uncertainty on eclipse fitting. Additionally, Visit 1 has a higher rate of spectral drift; while we perform corrections for this drift, it remains an additional source of uncertainty. Given these factors, we feel confident that the uncertainty we measure accurately reflects the sources of uncertainty in the data. 
%-----------------------------------------------------------------------------%
\section{Discussion}
\label{disc} 

\begin{figure}[htbp]
\centering
{
\includegraphics[width=135mm]{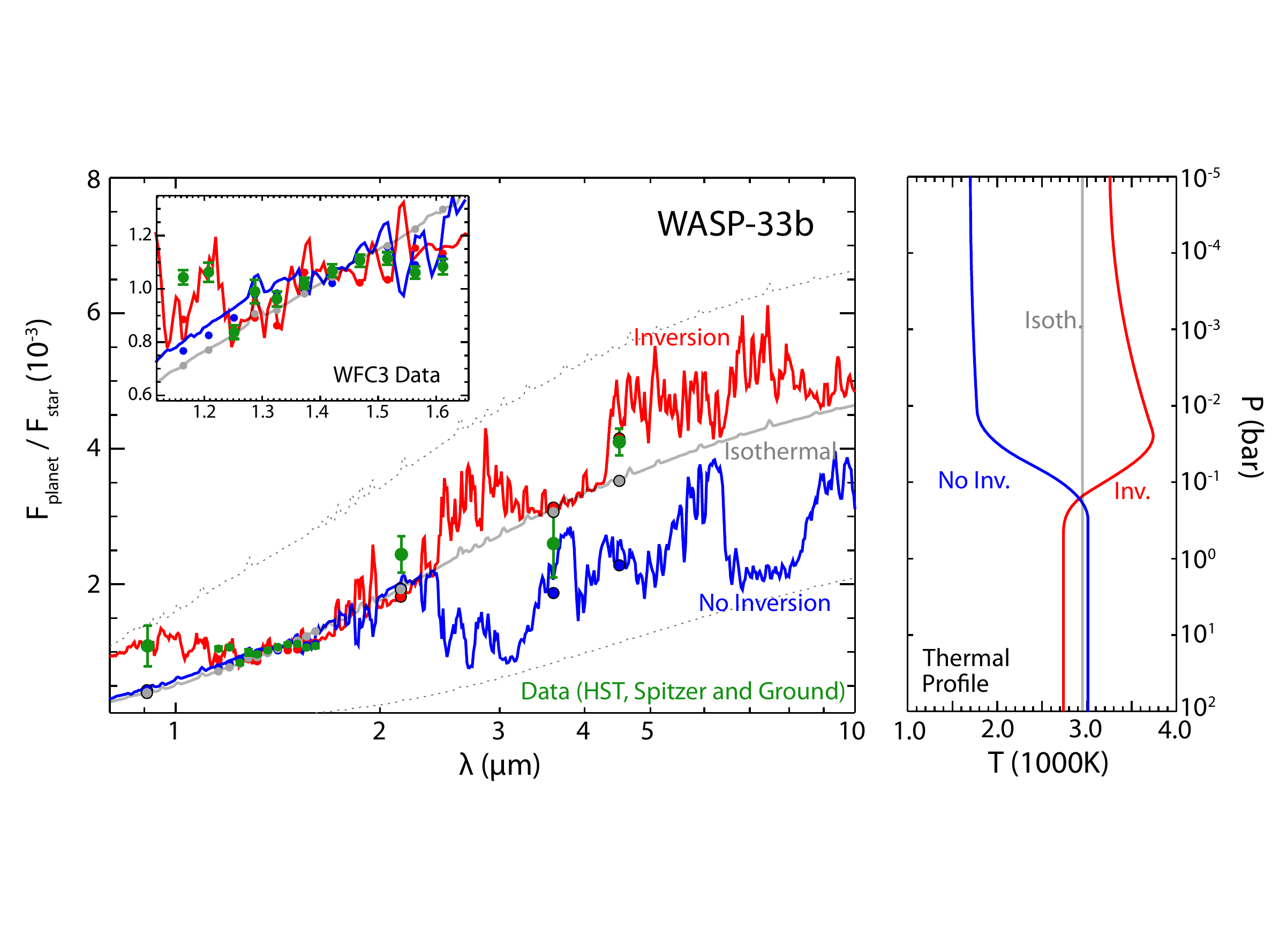}
\caption{The left panel shows the observed and model thermal emission spectra of WASP-33b. The observed WFC3 spectrum --- weighted mean of the two visits --- is shown in green in the 1.1-1.7 $\mu$m range, also shown in the inset. All the other photometric observations outside the WFC3 range from previous studies are also shown in green circles with error bars. The solid curves show three best-fit model spectra corresponding to three model scenarios: model with a thermal inversion (red), model without a thermal inversion (blue), model with an isothermal atmosphere (grey). The gray dotted lines show two blackbody spectra of the planet at temperatures of 1600 K and 3800 K. The corresponding colored circles show the models binned at the same resolution as the data. The red model, which corresponds to  an oxygen-rich atmosphere with a thermal inversion, provides the best fit to the data of all the models, as discussed in section~\ref{disc}. The right panel shows corresponding model pressure-temperature profiles. The best-fit model with (without) a thermal inversion is shown in red (blue), and the best-fit isothermal model is shown in grey. }
 \label{fig:p2madhu}
}
\end{figure} 

The hot Jupiter WASP-33b is one of the most irradiated hot Jupiters known and hence is among the most favorable candidates to host a thermal inversion in its dayside atmosphere. Studies in the past have suggested that extremely irradiated hot Jupiters should host thermal inversions due to strong absorption of incident stellar light by absorbers such as TiO and VO \citep{Hubeny2003, Fortney2008}. While \citet{Spiegel2009} have suggested that TiO and VO may not remain aloft in some hot Jupiter atmospheres due to downward drag by gravitational settling and condensation overtaking upward vertical mixing, the extreme irradiation of WASP-33b should maintain atmospheric temperatures above the TiO condensation point at all altitudes. However, alternate theories regarding the presence of thermal inversions do not depend solely on temperature. \citet{Madhusudhan2011a} and \citet{Madhusudhan2012a} suggested that high C/O ratios could also deplete inversion-causing compounds such as TiO and VO in hot Jupiters, thereby precluding the formation of thermal inversions, and \citet{Knutson2010a} proposed that the formation of inversions may instead be correlated with chromospheric activity, implying that hot Jupiters orbiting active stars are less likely to host thermal inversions (though their study did not include A-stars such as WASP-33). While the existence of an inversion has been questioned in the archetype planet HD 209458b \citep{Diamond-Lowe2014, Schwarz2015}, it is nevertheless reasonable to hypothesize that strong stellar irradiation may cause substantial perturbations in the temperature structure of hot Jupiter atmospheres. Given its extreme atmospheric conditions and bright thermal emission, WASP-33b presents a valuable opportunity to constrain the various hypotheses regarding thermal inversions in hot Jupiters, but previously reported photometric observations from Spitzer and ground-based facilities have been unable to conclusively constrain the presence of an inversion \citep{Deming2012, DeMooij2013}.  With the inclusion of our spectrum from the WFC3 instrument on HST, we can significantly improve these constraints. 

%\begin{figure}[htbp]
%\centering
%{
%\includegraphics[height=130mm]{figure6.eps}
%\caption{Model pressure-temperature profiles corresponding to the model spectra shown in Figure~\ref{fig:p2madhu}. The best-fit model with (without) a thermal inversion is shown in red (green), and the best-fit isothermal model is shown in brown.}
% \label{fig:pt-madhu}
%}
%\end{figure} 

\subsection{Atmospheric Models and Parameter Retrieval}
\label{models}
We modeled the temperature and composition of the planet's atmosphere and retrieved its properties using the retrieval technique of \citet{Madhusudhan2011a} and \citet{Madhusudhan2012a}. The model computes line-by-line radiative transfer for a plane-parallel atmosphere with the assumptions of hydrostatic equilibrium and global energy balance, as described in \citet{Madhusudhan2009}. The composition and pressure-temperature ($P$-$T$) profile of the dayside atmosphere are free parameters in the model. The model includes all the major opacity sources expected in hot Jupiter atmospheres, namely H$_2$O, CO, CH$_4$, CO$_2$, C$_2$H$_2$, HCN, TiO, VO, and collision-induced absorption (CIA) due to H$_2$-H$_2$, as described in \citet{Madhusudhan2012a}. Our molecular line lists are obtained from \citet{Freedman:08}, Freedman (personal communication, 2009), \citet{Rothman:05}, \citet{Karkoschka:10}, and Karkoschka (personal communication, 2011). Our CIA opacities are obtained from \citet{Borysow:97} and \citet{Borysow:02}. A Kurucz model \citet{Castelli:04} is used for the stellar spectrum, and the stellar and planetary parameters are adopted from \citet{Cameron2010}. 

We used our WFC3 observations together with previously published photometric data \citep{Deming2012, DeMooij2013, Smith2011} to obtain joint constraints on the chemical composition and temperature structure of the planet. We explored the model parameter space using a Markov Chain Monte Carlo algorithm \citep{Madhusudhan2011a} and determined regions of model space that best explain the data. Our model space includes models with and without thermal inversions, and models with oxygen-rich as well as carbon-rich compositions. To accommodate the uncertainty in the overall band offset resulting from our separate fits to the band-integrated light curve and each individual bin, we allowed a constant offset on the WFC3 spectrum as a free parameter in our model fits, with a prior constraint on the explored range based on the derived band-integrated uncertainty. Thus, the model has twelve free parameters: five for the $P$-$T$ profile, six for uniform mixing ratios of six molecules (H$_2$O, CO, CH$_4$, CO$_2$, C$_2$H$_2$, HCN), and one parameter for the WFC3 offset. For the inversion models we set the TiO and VO abundances to their solar abundance composition, whereas for the non-inversion model we assume TiO and VO are not present in significant quantities. We ran separate model fits to the data assuming inverted and non-inverted temperature profiles. We also investigated fits with isothermal temperature profiles which result in featureless black body spectra; such a model has only two free parameters, the isothermal temperature and the WFC3 offset.

We find that the sum-total of observations are best explained by a dayside atmosphere with a temperature inversion and an oxygen-rich composition with a slightly sub-solar abundance of H$_2$O (see Figure \ref{fig:p2madhu}). Previous photometric observations were consistent with two distinct models \citep{Deming2012}: (a) a model with oxygen-rich composition with a strong thermal inversion, and (b) a model with a carbon-rich composition but with no thermal inversion. In our current work, we use our WFC3 observations to break the degeneracies between these models, constrain the abundance of H$_2$O, and provide strong evidence for a temperature inversion caused by TiO. Figure~\ref{fig:p2madhu} shows the observed spectrum along with three best-fit model spectra in three model categories, one with a thermal inversion, one without a thermal inversion, and another with an isothermal profile. The corresponding pressure-temperature profiles are also shown. The best-fit inverted model has a $\chi^2$ of 98, the best-fit non-inverted model has a $\chi^2$ of 243, and the best-fit blackbody (BB) spectrum has a $\chi^2$ of 351. The causes of the remaining differences between the best-fit thermally-inverted atmosphere model and the WFC3 data points are unclear at this point, but the relative quality of fit between the three models can still be assessed robustly using the Bayesian Information Criterion (BIC) given by BIC$=\chi^2 + k\ln(N)$, where $k$ is the number of free parameters and $N$ is the number of data points (15). The BIC for the three best-fit models described above are 130.5 for the inverted model, 275.5 for the non-inverted model, and 356.4 for the blackbody model, implying that the inverted model provides a significantly better fit to all the data and that the spectrum is not a blackbody.  

We note that for each model category a population of `best-fit' models are found with similar $\chi^2$ values. Here we choose the most physically and chemically plausible model for each category by determining the most likely combinations of molecular mixing ratios in these solutions for the corresponding temperatures assuming equilibrium or non-equilibrium chemistry \citep{Madhusudhan2012a, Moses2013}. For example, the criteria used for O-rich models are that (a) CO must be comparable to the well-constrained H$_2$O abundance, (b) CH$_4$,C$_2$H$_2$, and HCN are below 10$^{-5}$, and (c) CO2 is below 10$^{-6}$. The best-fit inversion model has an O-rich composition with emission features due to CO, TiO, and H$_2$O. The mixing ratios of CO and H$_2$O in the best-fit model are marginally sub-solar at $\sim$$10^{-4}$ each, whereas TiO and all the other molecules (e.g. CH$_4$, C$_2$H$_2$, HCN) have nearly solar mixing ratios; i.e. consistent with mixing ratios predicted by chemical equilibrium assuming solar elemental abundances. The model fit to the 4.5 $\mu$m IRAC data point is due to the strong CO emission feature, whereas the TiO emission feature is responsible for the fit in the $z^\prime$ band (at 0.9 $\mu$m) and the bluer half of our WFC3 data. Low-amplitude H$_2$O emission features provide reasonable fits to the remaining data in the WFC3 bandpass, whereas the continuum emission is set by the temperature in the lower atmosphere of the inverted temperature profile shown in Figure~\ref{fig:p2madhu}. On the contrary, the non-inversion model fit shown in Figure~\ref{fig:p2madhu} has a C-rich composition, as discussed below, consistent with the findings of \citet{Deming2013}, and has a significantly poorer fit compared to the O-rich inversion model as discussed above. While the non-inversion model provides a very good fit to most of the WFC3 data, it provides a significantly poor fit to the two bluest WFC3 data points, the $z^\prime$ point, and the 4.5 $\mu$m IRAC point.

\begin{figure}[htbp]
\centering
{
\includegraphics[height=130mm]{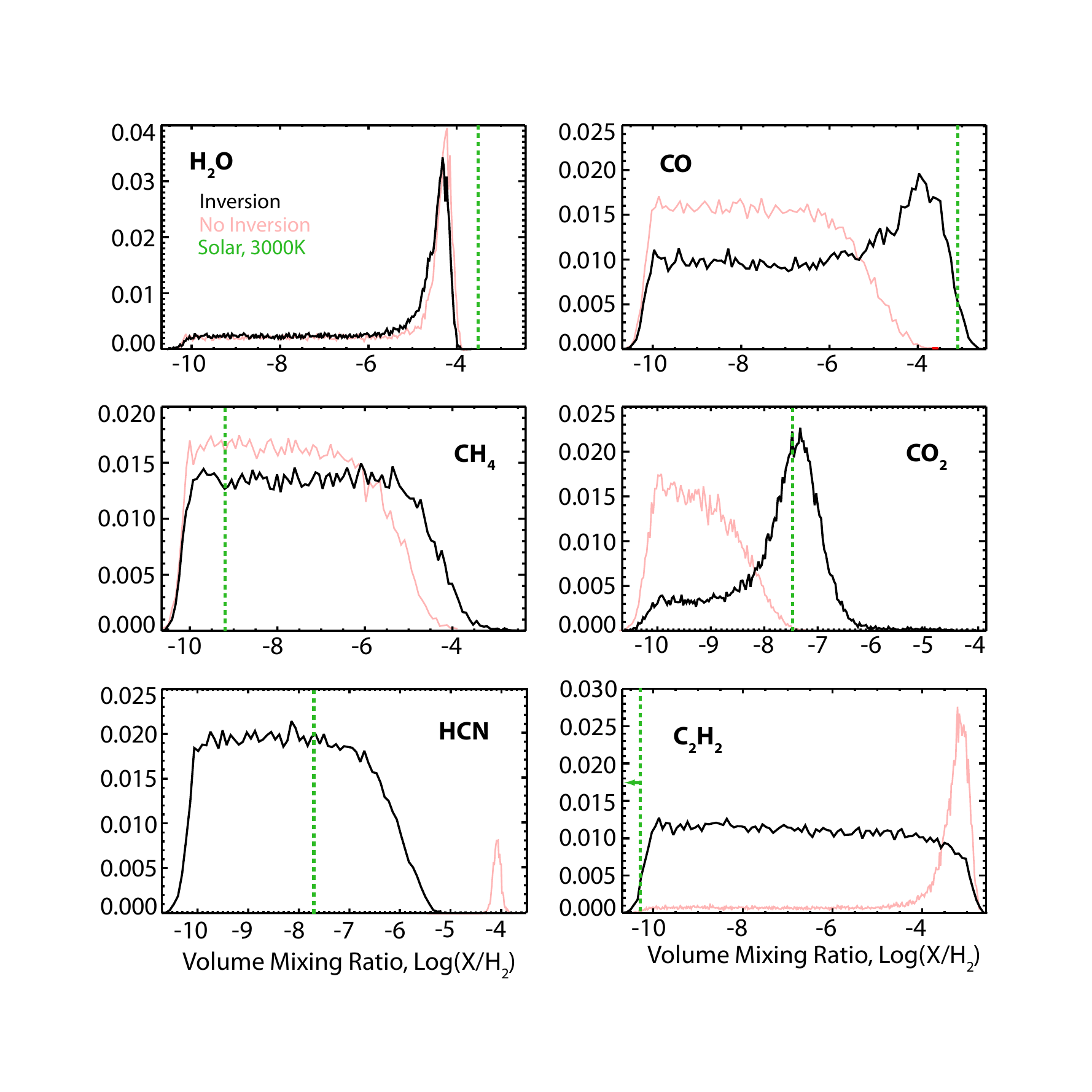}
\caption{Posterior probability distributions of the molecular mixing ratios for models with a thermal inversion (black) and without a thermal inversion (red), obtained by fitting the models to data using an MCMC retrieval method. Note that only the thermal inversion model provides a good fit to the data (as discussed in section~\ref{models}) and hence only the black posterior distributions represent meaningful constraints on the atmospheric composition. The non-inverted model does not provide good fits to the data but the corresponding posteriors in the compositions are shown here for completeness. For reference, mixing ratios predicted assuming solar elemental abundances and chemical equilibrium at 3000 K and 1 bar pressure are marked with green dashed lines. For the inversion model, the derived H$_2$O mixing ratio is relatively well constrained to values of 10$^{-5}$ - 10$^{-4}$ whereas only upper-limits are obtained for all the other molecules; H$_2$O and CO are found to be moderately sub-solar, while the other molecules are consistent with solar-abundance composition.}
 \label{fig:molposteriors}
}
\end{figure}

Figure~\ref{fig:molposteriors} shows the posterior probability distributions of the chemical compositions for each model, inverted versus non-inverted. Note that even though the non-inverted model provides much worse fits to the data than the inverted model, we show the posterior distributions on compositions for both models for completeness. The posteriors for the inverted model are consistent with an O-rich atmosphere, albeit of marginally sub-solar metallicity. The H$_2$O abundance is well constrained to between 10$^{-5}$ - 10$^{-4}$, thanks to the WFC3 bandpass which overlaps with a strong H$_2$O band, whereas only upper-limits are obtained for all the other molecules. The CO and H$_2$O abundances are marginally sub-solar, but the upper-limits on all the remaining molecules are consistent with a solar-type O-rich abundance pattern. On the other hand, the posteriors for the non-inverted model similarly constrain the H$_2$O abundance but also require high abundances of HCN and C$_2$H$_2$ which are possible only if the atmosphere is carbon-rich (i.e. C/O $\geq$ 1; Madhusudhan 2012; Kopparapu et al. 2013; Moses et al. 2013). However, these non-inverted models provide much worse fits to the data compared to the inverted models, as discussed above, and hence the corresponding constraints on the compositions are irrelevant. Nevertheless, these results still demonstrate that it is in principle possible to detect C-rich atmospheres using a combination of WFC3 and Spitzer data.

\begin{figure}[htbp]
\centering
{
\includegraphics[width=135mm]{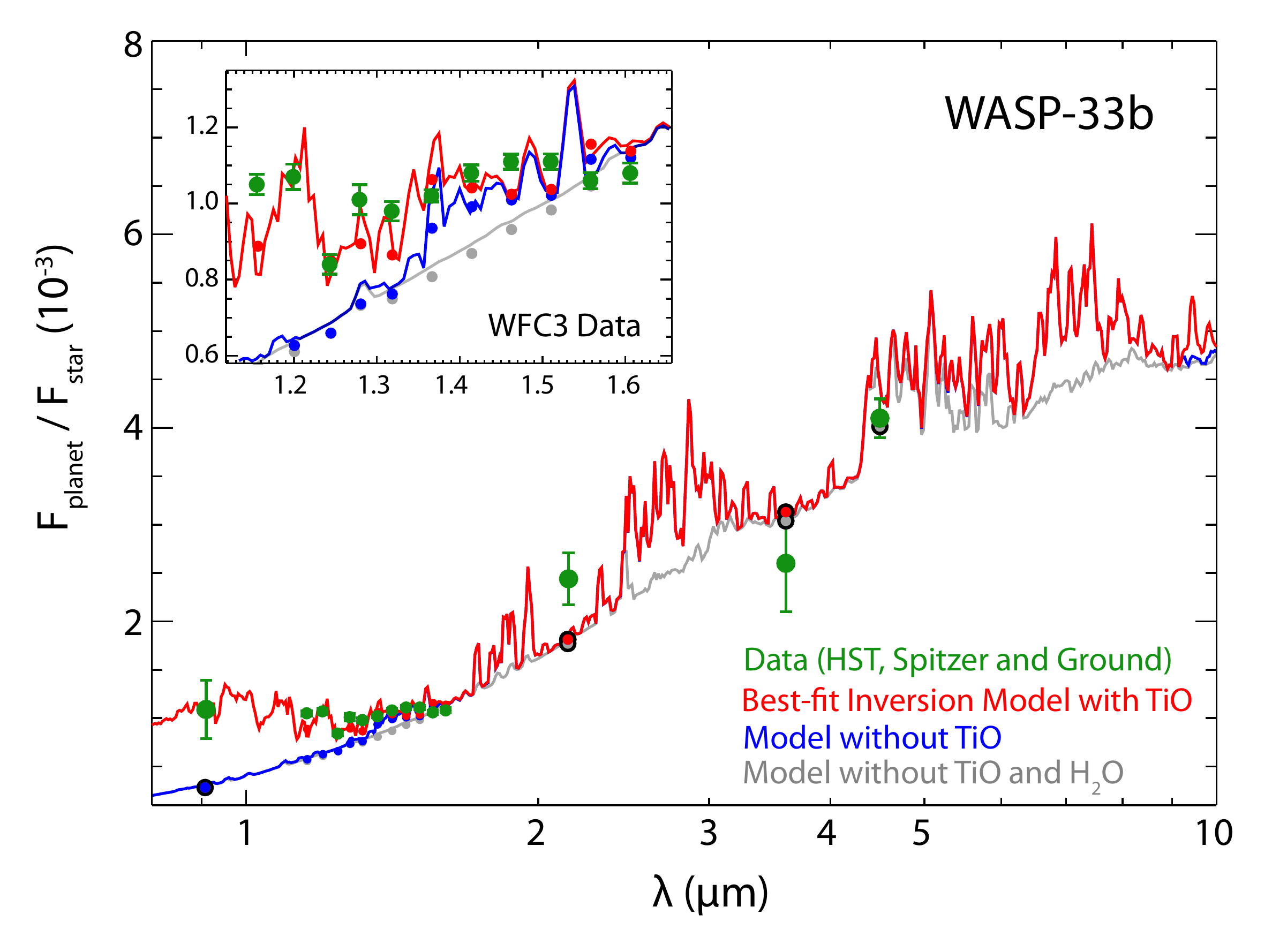}
\caption{Observed and model thermal emission spectra of WASP-33b, showing the effect of the inclusion of water and TiO on model spectra that include a temperature inversion. The observed WFC3 spectrum (green) and best-fit model (red) are as described in Figure \ref{fig:p2madhu}, where the red line indicates an inverted model including TiO and water. The remaining model curves show the effect of removing molecules from the model spectra: an inverted model without contributions from TiO (blue), and an inverted model without TiO or water (gray). It is clear that the presence of TiO is required to achieve a good fit.}
 \label{fig:p2madhuTiO}
}
\end{figure} 

The inversion model fit supports not just evidence for a temperature inversion, but also argues that the temperature inversion is due to TiO. Figure~\ref{fig:p2madhuTiO} shows the observed spectrum along with three model spectra, all including a thermal inversion: one is our best-fit model including both TiO and water, one lacks TiO, and the third lacks both TiO and water. It is clear that not only a temperature inversion, but also the presence of TiO (and water) is required in order to achieve a truly good fit to the observed data. As shown in previous studies \citep{Hubeny2003, Fortney2008}, strong absorption of incident light due to the strong UV/visible opacity of TiO can cause thermal inversions in hot Jupiters. On the other hand, the infrared opacity of TiO contributes to the emission features of TiO in the emergent spectrum of the planet similar to the emission features of other molecules in the planetary atmosphere caused by a thermal inversion. Thus the simultaneous inference of a thermal inversion and the presence of TiO presents a self-consistent case in favor of the results. Our inference of TiO supports previous theoretical predictions that TiO should be abundant in the hottest of oxygen-rich hot Jupiters, due to the lack of an effective vertical or day-night cold trap \citep{PerezBecker:2013hk, Parmentier:2013fd}. However, previous searches for TiO in other hot Jupiters using transmission spectroscopy have yielded either secure non-detections \citep{Huitson:2013gh,Sing2013} or inconclusive results \citep{Desert2008}. Two of these planets (HD 209458b and WASP-19b) are significantly cooler than WASP-33b, but the lack of TiO absorption in the transmission spectrum of WASP-12b (T$_{eq}\sim2500$) may instead be due to either a high-altitude haze layer obscuring any molecular absorption \citep{Sing2013} or a chemical composition that is carbon-rich rather than oxygen-rich \citep{Madhusudhan2011a, Stevenson:2014fo}.

However, since we cannot resolve individual spectral bands or lines of TiO in our spectrum, the evidence for TiO emission towards the blue end of WFC3 and in the $z^\prime$ band is not conclusive. Ostensibly, the presence of hazes/clouds in the atmosphere could lead to significant particulate scattering at short wavelengths, as the scattering cross-sections typically scale as an inverse-power-law of the wavelength (e.g. \citealp{Evans2013}; \citealp{Sing2013}). However, such an interpretation is met with several challenges. Firstly, the blue-ward flux would need to start rising abruptly below $\sim$1.25 $\mu$m, following a 20\% increase in reflected light over a 0.05$\mu$m spectral bin, which is inconsistent with the typical power law slope expected for the scattered light spectrum. Secondly, the planet orbits an A star ($T \sim 7400$ K) for which the spectrum peaks at relatively short wavelengths (0.43 $\mu$m). Consequently, the dominant contribution of the reflected light would be expected to be in the far blue with significantly less contribution in the near-infrared which is where the current data need strong flux.  Therefore the contribution from emission by TiO represents the most plausible explanation for the rise in the data at short wavelengths, but this inference can be further verified by future observations using existing facilities. TiO has strong spectral features in the red optical and near-infrared, between $\sim$0.7-1.1 $\mu$m as shown in Figure~\ref{fig:p2madhuTiO} (also see e.g. \citealp{Fortney2008}; \citealp{Madhusudhan2012a}). Several existing instruments can enable observations in this bandpass, including HST WFC3 G102 grism spectroscopy in the $\sim$ 0.8-1.15 $\mu$m range (Sing et al. 2014), and ground-based spectroscopy and/or photometry in the $\sim$0.7-1.2 $\mu$m range (e.g. \citealp{Bean2011}; \citealp{Fohring2013}; \citealp{Chen2014}).

Finally, we note that previous studies, both theoretical and observational, have suggested that the hottest exoplanets may be the most inefficient at redistributing heat to their night sides \citep{Cowan:2011kw,PerezBecker:2013hk}. Our results are consistent with those findings; if we compare the incoming radiation from the star with the outgoing day-side flux from our best-fit inverted model, we derive a low day-night redistribution ($\lesssim 15\%$), as would be expected for a planet with day-side temperatures above 2200K.  In contrast, the best-fit non-inverted model, which provides a poorer fit to the data compared to the inverted model, has very efficient redistribution ($\lesssim 50\%$).

%-----------------------------------------------------------------------------%
\section{Conclusion}
\label{Con}

In this paper we present our analysis of WFC3 observations of two occultations of WASP-33b, a hot Jupiter orbiting a \ds star. We reduce and analyze the spectroscopic time series for both visits, and correct for stellar oscillations of the star, as well as for motion of the target on the detector. We bin our spectrum, and achieve an RMS $\sim$1.05 times the photon noise. We compare our final emission spectrum to atmospheric models testing a range of carbon to oxygen ratios and temperature profiles, and find strong evidence for an oxygen-rich atmosphere that hosts a temperature inversion. We also present the first observational evidence for TiO in the dayside of an exoplanet atmosphere. This is consistent with, and improves upon, previous observations that could not discern between competing models, and demonstrates the power of combining HST, {\it Spitzer} and ground-based observations to break degeneracies in the composition and temperature structure of extrasolar planets.  Future measurements for a larger sample of exoplanets will help to determine the conditions under which thermal inversions exist, and pave the way for more detailed investigations with future instruments such as the James Webb Space Telescope.

%-----------------------------------------------------------------------------%
\section{Acknowledgements}

The authors would like to thank the anonymous referee for thoughtful comments that improved the paper. This work is based on observations made with the NASA/ESA Hubble Space Telescope that were obtained at the Space Telescope Science Institute, which is operated by the Association of Universities for Research in Astronomy, Inc., under NASA contract NAS 5-26555. These observations are associated with program GO-12495. Support for this work was provided by NASA through a grant from the Space Telescope Science Institute, with additional support for data analysis provided by a grant from the NASA Astrophysics Data Analysis Program (for K.H. and A.M.M.).
\clearpage
\bibliographystyle{apj}

%-----------------------------------------------------------------------------%
\end{document}